\documentclass[twocolumn,showpacs,preprintnumbers,amsmath,amssymb,prb,superscriptaddress]{revtex4}

\usepackage[dvips]{graphicx}
\usepackage{dcolumn}
\usepackage{bm}
\usepackage{mathrsfs}

\begin{document}
\title{High-temperature terahertz optical diode effect without magnetic order in polar FeZnMo$_3$O$_8$}
\author{Shukai Yu}
 \affiliation{Department of Physics and Engineering Physics, Tulane University, 6400 Freret St., New Orleans, LA 70118, USA}
\author{Bin Gao}
 \affiliation{Rutgers Center for Emergent Materials and Department of Physics and Astronomy, Piscataway, NJ 08854, USA}
\author{Jae Wook Kim}
 \affiliation{Rutgers Center for Emergent Materials and Department of Physics and Astronomy, Piscataway, NJ 08854, USA}
\author{Sang-Wook Cheong}
 \affiliation{Rutgers Center for Emergent Materials and Department of Physics and Astronomy, Piscataway, NJ 08854, USA}
\author{Michael K.L. Man}
 \affiliation{Femtosecond Spectroscopy Unit, Okinawa Institute of Science and Technology Graduate University}
\author{Julien Mad\'eo}
 \affiliation{Femtosecond Spectroscopy Unit, Okinawa Institute of Science and Technology Graduate University}
\author{Keshav M. Dani}
 \affiliation{Femtosecond Spectroscopy Unit, Okinawa Institute of Science and Technology Graduate University}
\author{Diyar Talbayev}
 \email{dtalbayev@gmail.com}
 \affiliation{Department of Physics and Engineering Physics, Tulane University, 6400 Freret St., New Orleans, LA 70118, USA}

\date{\today}

\newcommand{\fzmo}{FeZnMo$_3$O$_8$}
\newcommand{\cm}{\:\mathrm{cm}^{-1}}
\newcommand{\T}{\:\mathrm{T}}
\newcommand{\mc}{\:\mu\mathrm{m}}
\newcommand{\ve}{\varepsilon}
\newcommand{\dg}{^\mathtt{o}}

\begin{abstract}
We present a terahertz spectroscopic study of polar ferrimagnet FeZnMo$_3$O$_8$.  Our main finding is a giant high-temperature optical diode effect, or nonreciprocal directional dichroism, where the transmitted light intensity in one direction is over 100 times lower than intensity transmitted in the opposite direction.  The effect takes place in the paramagnetic phase with no long-range magnetic order in the crystal, which  contrasts sharply with all existing reports of the terahertz optical diode effect in other magnetoelectric materials, where the long-range magnetic ordering is a necessary prerequisite.  In \fzmo, the effect occurs resonantly with a strong magnetic dipole active transition centered at 1.27 THz and assigned as electron spin resonance between the eigenstates of the single-ion anisotropy Hamiltonian.  We propose that the optical diode effect in paramagnetic FeZnMo$_3$O$_8$ is driven by signle-ion terms in magnetoelectric free energy.
\end{abstract}

\maketitle


Multiferroic materials that combine ferroelectricity with magnetism have been the source of fascinating physical phenomena and functionalities\cite{cheong:13, ramesh:21}.  One example is the terahertz (THz) nonreciprocal directional dichroism that was recently discovered in multiferroics\cite{kezsmarki:057403, bordacs:734}.  The term refers to the difference in absorption coefficient for linearly polarized light waves traveling in opposite directions.  A material can be transparent for light traveling in one direction and completely opaque when the same light wave travels in the opposite direction.  By analogy with the semiconductor diode, we will use the term optical diode effect (ODE) for the directional dichroism.

\begin{figure}[ht]
\begin{center}
\includegraphics[width=2.5in]{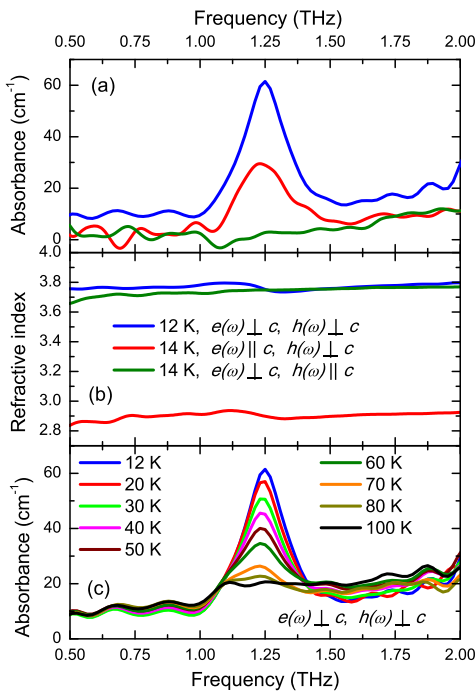}
\caption{\label{fig:fig1}(Color online) (a,b) Absorption and refractive index in zero applied magnetic field ($\bm{H}=0$) for different polarizations of the incident THz wave.  (c) Temperature dependence of absorption.}
\end{center}
\end{figure}

ODE functionality can potentially find applications in THz optical isolators.  Practical ODE devices would need to display the effect close to room temperature, while most of the reported ODE observations until now occured in low-temperature magnetically ordered states\cite{kezsmarki:057403, takahashi:121, bordacs:734, kibayashi:4583, takahashi:037204, bordacs:214441, kuzmenko:174407, kuzmenko:184409, takahashi:180404, narita:094433, masuda:041117}.  ODE exists at room temperature in ferroelectric antiferromagnet BiFeO$_3$, but its magnitude is rather small\cite{kezsmarki:127203}.  In this letter, we report the observation of a giant high-temperature THz ODE in polar ferrimagnet \fzmo.  We demonstrate complete suppression of absorption for one direction of traveling light, while the absorption for the other direction remains very strong, resulting in 100-fold difference in transmitted light intensity between the two directions.  Most remarkably, the strong ODE persists in the high-temperature paramagnetic state of \fzmo\: without long-range magnetic order up to 110 K, which is in stark contrast with all previous reports of THz ODE where the presence of magnetic order is a necessity.  The ODE and related optical magnetoelectric effects in room-temperature paramagnetic state were first reported by Rikken et al. at visible wavelengths, were the strength of the effect was many orders of magnitude smaller\cite{rikken:133005, roth:4478, roth:063001}. The ODE in \fzmo is resonant with a 1.27 THz electron spin transition between the eigenstates of single-ion anisotropy Hamiltonian, which is the dominant energy scale.  We suggest that the high-temperature giant ODE in \fzmo\: results from single-ion terms in magnetoelectric interaction, which can open a new direction in the quest for stronger ODE in other magnetoelectrics.

High quality monodomain crystals of \fzmo\: were prepared by chemical vapor transport method\cite{wang:12268}.  Two platelet-shaped crystals were selected, one with the polar $c$ axis perpendicular to the platelet, the other with the $c$ axis in the platelet plane.  THz time domain spectroscopy (TDS) in high magnetic field (up to 16 T) was used to study the low-energy excitations\cite{silwal:092116}.

\begin{figure}[ht]
\begin{center}
\includegraphics[width=2.5in]{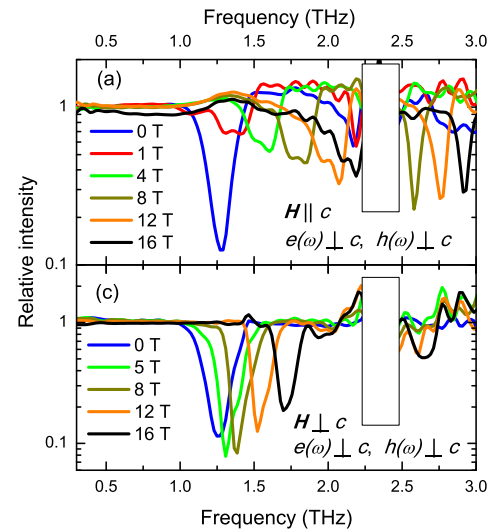}
\caption{\label{fig:fig2}(Color online) Magnetic field dependence of transmitted intensity at $T=4$ K.  (a) Magnetic field $\bm{H}$ applied along the $c$ axis in Faraday measurement geometry.  (b) Magnetic field $\bm{H}$ applied perpendicular to the $c$ axis in Voigt measurement geometry.}
\end{center}
\end{figure}

\fzmo\: is derived from the parent material Fe$_2$Mo$_3$O$_8$ by susbtituting Fe with Zn.  Both materials adopt a polar space group $P$6$_3mc$ \cite{mccarroll:5410,wang:12268,kurumaji:031034,kurumaji:020405}.  In the parent, Fe ions occupy octahedrally and tetrahedrally oxygen coordinated sites, while Zn ions preferentially populate the tetrahedral sites in the substituted variant\cite{kurumaji:031034}.  With equal amounts of Fe and Zn in \fzmo, exchange interactions are heavily diluted.  No signature of long range magnetic order is found down to $T_C=14$ K\cite{bertrand:379}, where the ferrimagnetic ground state appears with antiparallel alignment of spins on octahedral and tetrahedral sites.  $M(H)$ curves along the $c$ axis show a very narrow hysteresis loop with coercivity $H_C=0.06$ T and saturation magnetic moment of 3.9 $\mu_B$/f.u.  The $M(H)$ measurement perpendicular to the $c$ axis does not show saturation up to 7 T, indicating a very strong magnetic anisotropy\cite{sm:odefzmo}.

Figures~\ref{fig:fig1}(a,b) show the recorded absorption and refractive index spectra at low temperature (12 K and 14 K) for different polarizations of the incident THz wave.  Polarization analysis shows that the 1.27 THz resonance is magnetic-dipole active, as the absorption disappears completely when the THz magnetic field $h(\omega)$ is oriented along the polar $c$ direction.  Cooling the crystal to 4 K into the ferrimagnetic state does not appreciably change the THz absorption spectra.  A similar resonance near 1.4 THz was observed by Kurimaji et al.\cite{kurumaji:077206} in (Fe$_{0.6}$Zn$_{0.4}$)$_2$Mo$_3$O$_8$, along with weaker resonances at 2.3 and 2.5 THz.  The weak 2.3 and 2.5 THz resonances are outside the frequency window of our zero-field THz TDS spectrometer.  Kurumaji et al. suggest that the 1.4 THz mode in (Fe$_{0.6}$Zn$_{0.4}$)$_2$Mo$_3$O$_8$ may be related to local single-site spin transitions on Fe sites\cite{kurumaji:077206}.  We show that the 1.27 THz resonance in \fzmo\: is indeed an electron spin resonance between the eigenstates of the single-ion anisotropy Hamiltonian on Fe$^{2+}$ cites.   

Figure~\ref{fig:fig1}(c) shows the temperature dependence of the 1.27 THz absorption, which persists almost up to 100 K temperature.  The strong absorption in the paramagnetic state serves evidence that the magnetic-dipole transition at 1.27 THz cannot be related to a collective excitation associated with an ordered magnetic state.  Figure~\ref{fig:fig2} shows the magnetic field dependence of the THz transmission spectra with oscillating THz fields $e(\omega)$ and $h(\omega)$ both perpendicular to the $c$ axis.  The static magnetic field $\bm{H}$ is applied both along and perpendicular to the $c$ axis.  The frequency window of the high-magnetic-field THz TDS setup includes frequencies up to 3 THz, albeit with a narrow "dark" band between 2.2-2.5 THz due to absorption in polyethylene lenses that are part of the setup.  This dark band is blanked out in Fig~\ref{fig:fig2}.  In high magnetic field, our setup does not allow the collection of a free space reference spectrum, instead of which we used a reference spectrum that is an average of spectra at different recorded magnetic fields\cite{mihaly:024414}.  Due to the strong shift of the 1.27 resonance, such reference allows us to isolate and measure only the magnetic field-dependent absorption.  In Fig.~\ref{fig:fig2}, the relative transmission is the THz intensity at a fixed magnetic field divided by the intensity of above-defined reference spectrum.  Clearly, the position of the 1.27 THz resonance shifts with magnetic field, and the shift depends on the direction of applied field $\bm{H}$.  At high $\bm{H}$, we also find and additional field-dependent absorption in the 2.5-3.0 THz range (Fig.~\ref{fig:fig2}).

\begin{figure}[ht]
\begin{center}
\includegraphics[width=2.5in]{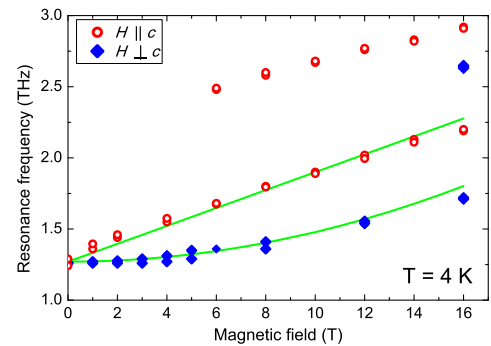}
\caption{\label{fig:freqvsb}(Color online) Magnetic field dependence of resonance frequencies at $T=4$ K.}  
\end{center}
\end{figure}

Figure~\ref{fig:freqvsb} summarizes the measured magnetic field dependence of the observed resonance frequencies.  The main 1.27 THz resonance shifts up linearly when $\bm{H}$ is applied along the $c$ direction; the resonance shifts up quadratically when $\bm{H}$ is applied perpendicular to $c$.  This behavior can be described well by the single-ion Hamiltonian
\begin{equation}
{\cal H}=-DS{_z}^2 + g \mu_B \bm{H}\cdot \bm{S},
\end{equation}
where $S=2$ is the spin on Fe$^{2+}$ ions, $D>0$ is the easy axis anisotropy constant, and $\bm{H}$ is the applied magnetic field.  In zero field, the oscillating THz field $h(\omega)$ drives the transitions between the eigenstates of the unperturbed part ${\cal H}=-DS{_z}^2$ with energy $\hbar\omega = 3D$. For magnetic field applied along $z$ and $x$ directions (parallel and perpendicular to the $c$ axis, respectively), we expect the following resonance frequency shifts
\begin{eqnarray}
\label{shiftz}
\hbar\omega = 3D + g\mu_BH_z,\\
\label{shiftx}
\hbar\omega = 3D + \frac{2(g\mu_BH_x)^2}{3D},
\end{eqnarray}
which were computed under the assumptions of an isotropic $g$ factor and of small frequency shifts compared to $3D$.  The solid green lines in Fig.~\ref{fig:freqvsb} are frequencies computed using Eqs. (\ref{shiftz}) and (\ref{shiftx}) with $g=4.50$.  The agreement with data is very good.  The measured temperature and magnetic field dependence of the 1.27 THz mode lead to the conclusion that it arises from the electron spin resonance transitions between the eigenstates of single-ion Hamiltonian (1).  In the paramagnetic state, the transisions occur on individual Fe ions.  In the ferrimagnetic state, the transitions correspond to the precession of the macroscopic magnetization of the majority Fe spins on octahderal sites with single-ion anisotropy as the dominant energy scale.  

The high-energy transitions seen in Figs.~\ref{fig:fig2} and~\ref{fig:freqvsb} above 2.5 THz are likely similar in origin to the 2.3 and 2.5 THz transitions observed in (Fe$_{0.6}$Zn$_{0.4}$)$_2$Mo$_3$O$_8$\cite{kurumaji:077206}.

We now focus on the ODE associated with the 1.27 THz resonance in \fzmo.  Two distinct mechanisms have been established for ODE.  One is magnetochiral dichroism found in magnetic and chiral materials\cite{bordacs:734,kezsmarki:3203}.  The other is toroidal dichroism, for which the material must possess simultaneous electric polarization $\bm{P}$ and magnetization $\bm{M}$\cite{kezsmarki:057403, miyahara:073708, miyahara:195145}.  The ODE happens for light propagating along and opposite the toroidal vector $\bm{T}=\bm{P} \times \bm{M}$\cite{sm:odefzmo}.  The THz wave must be polarized with its electric field $e(\omega)$ along $\bm{P}$ and with its magnetic field $h(\omega)$ along magnetization $\bm{M}$ while propagating along $\bm{T}$.  

\begin{figure}[ht]
\begin{center}
\includegraphics[width=2.5in]{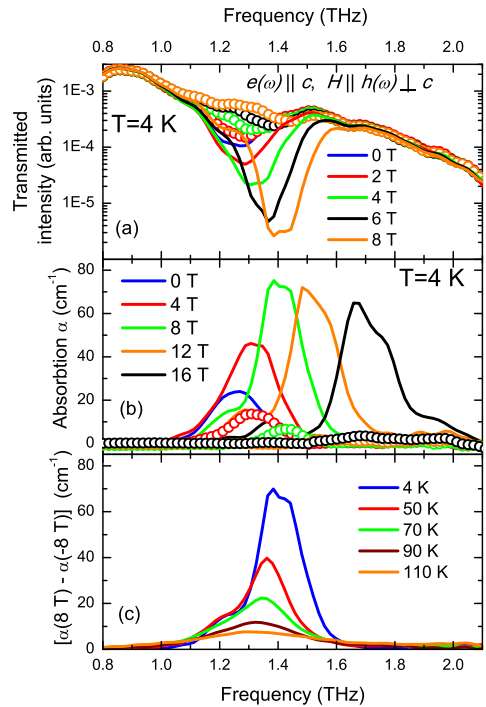}
\caption{\label{fig:fig4}(Color online) (a,b) Optical diode effect and absorption coefficient for positive and negative magnetic field $\bm{H}$. Solid lines show absorption coefficient for positive magnetic field.  Open circles of the same color show the absorption coefficient for negative magnetic field. (c) Temperature dependence of the optical diode effect.}
\end{center}
\end{figure}

 In \fzmo, the polarization $\bm{P}$ is along the $c$ axis.  Figure~\ref{fig:fig4}(a) demostrates the ODE in \fzmo\: in magnetic field $\bm{H}$ applied perpendicular to the $c$ axis at 4 K.  The THz wave is polarized along the $c$ axis ($e(\omega)\|\bm{P}$) and is traveling along the $\bm{P}\times\bm{H}$ direction.  ODE is detected by reversing the direction of $\bm{H}$.  Figure~\ref{fig:fig4}(a) displays the intensity of transmitted THz wave at the resonance frequency for positive and negative magnetic fields.  For both positive and negative fields, the resonance frequency shifts according to Eq. (\ref{shiftx}).   However, together with the frequency shift we also find that the intensity of the resonance changes.  For positive fields, the resonance is enhanced, for negative fields, the resonance is suppressed.  In negative 8 T field, the resonant absorption practically disappears; the difference in transmitted intensity between positive and negative 8 T field reaches a factor of 100.  Very clearly, this is a demonstration of a giant ODE.  

To quantitatively compare ODE in \fzmo\: to other materials, we compute the absorption coefficient for positive and negative magnetic fields, Fig.~\ref{fig:fig4}(b).  The difference in absorption coefficient reaches $\Delta\alpha=72$ cm$^{-1}$ between positive and negative magnetic fields in the 8-12 T range.  This is significantly higher than most reported values of ODE\cite{kezsmarki:057403, takahashi:121, bordacs:734, kibayashi:4583, bordacs:214441, kezsmarki:127203, kuzmenko:174407, kuzmenko:184409, takahashi:180404, narita:094433, masuda:041117}, with the only higher value $\Delta\alpha\sim400$ cm$^{-1}$ cited for Gd$_{0.5}$Tb$_{0.5}$MnO$_3$\cite{takahashi:037204}.  Figure~\ref{fig:fig4}(c) shows the temperature dependence of $\Delta\alpha$ for $\bm{H}=\pm8$ T.  ODE and $\Delta\alpha$ remain significant up to 110 K in the paramagnetic state of \fzmo, when no long range magnetic order is present in the crystal.  This observation drastically contrasts with all previous reports, where THz ODE occurs in a magnetically ordered state.  In the room-temperature paramagnetic state, the ODE was first observed by Rikken et al. at visible wavelengths\cite{rikken:133005}.  To quantitatively compare our THz ODE with the observations of Rikken, we compute the quantity $(\alpha(H)-\alpha(-H))/(\alpha(H)+\alpha(-H))/2H$ and find that it exceeds $5\times10^{-2}$ T$^{-1}$ in \fzmo\: at 4 K temperature.  In the work of Rikken et al., this quantity was measured to be $2.5\times10^{-5}$ T$^{-1}$ with $\Delta\alpha=1.2\times10^{-4}$ cm$^{-1}$.  By both measures, our THz ODE is many orders of magnitude stronger.

What is the origin of the high-temperature ODE in \fzmo?  In addition to the presence of the toroidal moment $\bm{T}=\bm{P} \times \bm{M}$ in the crystal, another prerequisite for ODE is nonzero dynamic magnetoelectric susceptibility $\chi_{xz}^{me}(\omega)$\cite{miyahara:073708, miyahara:195145} such that
\begin{eqnarray}
\mathtt{Im}\: \chi_{xz}^{me}(\omega) = \sum_n \frac{\pi c\mu_0}{2\hbar NV}(\langle0\vert h^x \vert n\rangle\langle n\vert e^z\vert 0\rangle +\nonumber\\
 \langle0\vert e^z \vert n\rangle\langle n\vert h^x\vert 0\rangle) \delta(\omega-\omega_n).
\end{eqnarray}
Here, $\langle0\vert h^x \vert n\rangle$ and $\langle0\vert e^z \vert n\rangle$ are magnetic and electric dipole matrix elements between ground and excited spin states of the Fe$^{2+}$ ions.  The first, magnetic dipole matrix elements is clearly not zero, as evidenced by the strong magnetic absorption when $h(\omega)$ is perpendicular to the $c$ axis.  We must conclude the existence of the second, electric dipole matrix element, from the observed strong ODE.  We propose that such electric dipole matrix elements could exist on tetrahedrally coordinated Fe$^{2+}$ sites with no inversion symmetry due to spin dependent metal-ligand hybridization\cite{jia:224444, arima:073702}.  Such interaction could induce an electric dipole at the spin site, as well as electric-field driven spin transitions.  The spin dependent metal-ligand hybridization was invoked to explain the ODE in Ba$_2$CoGe$_2$O$_7$ associated with the spin resonance due to anisotropy gap excitation\cite{miyahara:073708}.  In contrast to our present results, the ODE in Ba$_2$CoGe$_2$O$_7$ disappeared above the magnetic ordering temperature $T_N=6.7$ K\cite{kezsmarki:057403}.  Further work is needed to clarify the details that govern the magnetoelectric susceptibility $\chi_{xz}^{me}(\omega)$.  For example, in \fzmo\: composition with equal amount of Fe and Zn, where the tetrahedral sites are preferentially occupied by Zn and the octahedral sites are preferentially occupied by Fe, the 1.27 THz resonance would need to get some of its strength from tetrahedral sites to give rise to $\chi_{xz}^{me}(\omega)$.  Alternatively, octahedral sites may need to experience local inversion symmetry breaking to allow the same.

To summarize, we have reported a giant high-temperature ODE (110 K) in polar paramagnetic \fzmo\: without long-range magnetic order, which is fundamentally different all from prior reports of THz ODE.  The ODE in \fzmo\: happens at the frequency of the strong electron spin resonance assigned as the single-ion anisotropy gap excitation.  We have proposed that necessary dynamic magnetoelectric susceptibility $\chi_{xz}^{me}(\omega)$ also results from single-ion magnetoelectric interactions\cite{jia:224444, arima:073702}.  Our experimental results demonstrate that single-site magnetic and magnetoelectric interactions can provide a new avenue in the search for high-temperature THz ODE in other magnetoelectric materials.  

We acknowledge fruitful discussions with Valery Kiryukhin and Andrei Sirenko.  The work at Tulane University was supported by the NSF Award No. DMR-1554866.  The work at Rutgers University was supported by the DOE under Grant No. DOE: DE-FG02-07ER46382.


\begin{thebibliography}{31}
\expandafter\ifx\csname natexlab\endcsname\relax\def\natexlab#1{#1}\fi
\expandafter\ifx\csname bibnamefont\endcsname\relax
  \def\bibnamefont#1{#1}\fi
\expandafter\ifx\csname bibfnamefont\endcsname\relax
  \def\bibfnamefont#1{#1}\fi
\expandafter\ifx\csname citenamefont\endcsname\relax
  \def\citenamefont#1{#1}\fi
\expandafter\ifx\csname url\endcsname\relax
  \def\url#1{\texttt{#1}}\fi
\expandafter\ifx\csname urlprefix\endcsname\relax\def\urlprefix{URL }\fi
\providecommand{\bibinfo}[2]{#2}
\providecommand{\eprint}[2][]{\url{#2}}

\bibitem[{\citenamefont{Cheong and Mostovoy}(2007)}]{cheong:13}
\bibinfo{author}{\bibfnamefont{S.-W.} \bibnamefont{Cheong}} \bibnamefont{and}
  \bibinfo{author}{\bibfnamefont{M.}~\bibnamefont{Mostovoy}},
  \bibinfo{journal}{Nature Mat.} \textbf{\bibinfo{volume}{6}},
  \bibinfo{pages}{13} (\bibinfo{year}{2007}).

\bibitem[{\citenamefont{Ramesh and Spaldin}(2007)}]{ramesh:21}
\bibinfo{author}{\bibfnamefont{R.}~\bibnamefont{Ramesh}} \bibnamefont{and}
  \bibinfo{author}{\bibfnamefont{N.}~\bibnamefont{Spaldin}},
  \bibinfo{journal}{Nature Mat.} \textbf{\bibinfo{volume}{6}},
  \bibinfo{pages}{21} (\bibinfo{year}{2007}).

\bibitem[{\citenamefont{K\'ezsm\'arki et~al.}(2011)\citenamefont{K\'ezsm\'arki,
  Kida, Murakawa, Bord\'acs, Onose, and Tokura}}]{kezsmarki:057403}
\bibinfo{author}{\bibfnamefont{I.}~\bibnamefont{K\'ezsm\'arki}},
  \bibinfo{author}{\bibfnamefont{N.}~\bibnamefont{Kida}},
  \bibinfo{author}{\bibfnamefont{H.}~\bibnamefont{Murakawa}},
  \bibinfo{author}{\bibfnamefont{S.}~\bibnamefont{Bord\'acs}},
  \bibinfo{author}{\bibfnamefont{Y.}~\bibnamefont{Onose}}, \bibnamefont{and}
  \bibinfo{author}{\bibfnamefont{Y.}~\bibnamefont{Tokura}},
  \bibinfo{journal}{Phys. Rev. Lett.} \textbf{\bibinfo{volume}{106}},
  \bibinfo{pages}{057403} (\bibinfo{year}{2011}),
  \urlprefix\url{https://link.aps.org/doi/10.1103/PhysRevLett.106.057403}.

\bibitem[{\citenamefont{Bordacs et~al.}(2014)\citenamefont{Bordacs,
  K\'ezsm\'arki, Seki, and Tokura}}]{bordacs:734}
\bibinfo{author}{\bibfnamefont{S.}~\bibnamefont{Bordacs}},
  \bibinfo{author}{\bibfnamefont{I.}~\bibnamefont{K\'ezsm\'arki}},
  \bibinfo{author}{\bibfnamefont{S.}~\bibnamefont{Seki}}, \bibnamefont{and}
  \bibinfo{author}{\bibfnamefont{Y.}~\bibnamefont{Tokura}},
  \bibinfo{journal}{Nat. Comm.} \textbf{\bibinfo{volume}{5}},
  \bibinfo{pages}{4583} (\bibinfo{year}{2014}),
  \urlprefix\url{http://dx.doi.org/10.1038/ncomms5583}.

\bibitem[{\citenamefont{Takahashi et~al.}(2011)\citenamefont{Takahashi,
  Shimano, Kaneko, Murakawa, and Tokura}}]{takahashi:121}
\bibinfo{author}{\bibfnamefont{Y.}~\bibnamefont{Takahashi}},
  \bibinfo{author}{\bibfnamefont{R.}~\bibnamefont{Shimano}},
  \bibinfo{author}{\bibfnamefont{Y.}~\bibnamefont{Kaneko}},
  \bibinfo{author}{\bibfnamefont{H.}~\bibnamefont{Murakawa}}, \bibnamefont{and}
  \bibinfo{author}{\bibfnamefont{Y.}~\bibnamefont{Tokura}},
  \bibinfo{journal}{Nat. Phys.} \textbf{\bibinfo{volume}{8}},
  \bibinfo{pages}{121} (\bibinfo{year}{2011}).

\bibitem[{\citenamefont{Kibayashi et~al.}(2012)\citenamefont{Kibayashi,
  Takahashi, Szaller, Demko, Kida, Murakawa, Onose, Shimano, R\~o\ om, Nagel
  et~al.}}]{kibayashi:4583}
\bibinfo{author}{\bibfnamefont{S.}~\bibnamefont{Kibayashi}},
  \bibinfo{author}{\bibfnamefont{Y.}~\bibnamefont{Takahashi}},
  \bibinfo{author}{\bibfnamefont{D.}~\bibnamefont{Szaller}},
  \bibinfo{author}{\bibfnamefont{L.}~\bibnamefont{Demko}},
  \bibinfo{author}{\bibfnamefont{N.}~\bibnamefont{Kida}},
  \bibinfo{author}{\bibfnamefont{H.}~\bibnamefont{Murakawa}},
  \bibinfo{author}{\bibfnamefont{Y.}~\bibnamefont{Onose}},
  \bibinfo{author}{\bibfnamefont{R.}~\bibnamefont{Shimano}},
  \bibinfo{author}{\bibfnamefont{T.}~\bibnamefont{R\~o\ om}},
  \bibinfo{author}{\bibfnamefont{U.}~\bibnamefont{Nagel}},
  \bibnamefont{et~al.}, \bibinfo{journal}{Nat. Phys.}
  \textbf{\bibinfo{volume}{8}}, \bibinfo{pages}{734} (\bibinfo{year}{2012}).

\bibitem[{\citenamefont{Takahashi et~al.}(2013)\citenamefont{Takahashi,
  Yamasaki, and Tokura}}]{takahashi:037204}
\bibinfo{author}{\bibfnamefont{Y.}~\bibnamefont{Takahashi}},
  \bibinfo{author}{\bibfnamefont{Y.}~\bibnamefont{Yamasaki}}, \bibnamefont{and}
  \bibinfo{author}{\bibfnamefont{Y.}~\bibnamefont{Tokura}},
  \bibinfo{journal}{Phys. Rev. Lett.} \textbf{\bibinfo{volume}{111}},
  \bibinfo{pages}{037204} (\bibinfo{year}{2013}),
  \urlprefix\url{https://link.aps.org/doi/10.1103/PhysRevLett.111.037204}.

\bibitem[{\citenamefont{Bord\'acs et~al.}(2015)\citenamefont{Bord\'acs, Kocsis,
  Tokunaga, Nagel, R\~o\ om, Takahashi, Taguchi, and Tokura}}]{bordacs:214441}
\bibinfo{author}{\bibfnamefont{S.}~\bibnamefont{Bord\'acs}},
  \bibinfo{author}{\bibfnamefont{V.}~\bibnamefont{Kocsis}},
  \bibinfo{author}{\bibfnamefont{Y.}~\bibnamefont{Tokunaga}},
  \bibinfo{author}{\bibfnamefont{U.}~\bibnamefont{Nagel}},
  \bibinfo{author}{\bibfnamefont{T.}~\bibnamefont{R\~o\ om}},
  \bibinfo{author}{\bibfnamefont{Y.}~\bibnamefont{Takahashi}},
  \bibinfo{author}{\bibfnamefont{Y.}~\bibnamefont{Taguchi}}, \bibnamefont{and}
  \bibinfo{author}{\bibfnamefont{Y.}~\bibnamefont{Tokura}},
  \bibinfo{journal}{Phys. Rev. B} \textbf{\bibinfo{volume}{92}},
  \bibinfo{pages}{214441} (\bibinfo{year}{2015}),
  \urlprefix\url{http://link.aps.org/doi/10.1103/PhysRevB.92.214441}.

\bibitem[{\citenamefont{Kuzmenko et~al.}(2014)\citenamefont{Kuzmenko, Shuvaev,
  Dziom, Pimenov, Schiebl, Mukhin, Ivanov, Bezmaternykh, and
  Pimenov}}]{kuzmenko:174407}
\bibinfo{author}{\bibfnamefont{A.~M.} \bibnamefont{Kuzmenko}},
  \bibinfo{author}{\bibfnamefont{A.}~\bibnamefont{Shuvaev}},
  \bibinfo{author}{\bibfnamefont{V.}~\bibnamefont{Dziom}},
  \bibinfo{author}{\bibfnamefont{A.}~\bibnamefont{Pimenov}},
  \bibinfo{author}{\bibfnamefont{M.}~\bibnamefont{Schiebl}},
  \bibinfo{author}{\bibfnamefont{A.~A.} \bibnamefont{Mukhin}},
  \bibinfo{author}{\bibfnamefont{V.~Y.} \bibnamefont{Ivanov}},
  \bibinfo{author}{\bibfnamefont{L.~N.} \bibnamefont{Bezmaternykh}},
  \bibnamefont{and} \bibinfo{author}{\bibfnamefont{A.}~\bibnamefont{Pimenov}},
  \bibinfo{journal}{Phys. Rev. B} \textbf{\bibinfo{volume}{89}},
  \bibinfo{pages}{174407} (\bibinfo{year}{2014}),
  \urlprefix\url{https://link.aps.org/doi/10.1103/PhysRevB.89.174407}.

\bibitem[{\citenamefont{Kuzmenko et~al.}(2015)\citenamefont{Kuzmenko, Dziom,
  Shuvaev, Pimenov, Schiebl, Mukhin, Ivanov, Gudim, Bezmaternykh, and
  Pimenov}}]{kuzmenko:184409}
\bibinfo{author}{\bibfnamefont{A.~M.} \bibnamefont{Kuzmenko}},
  \bibinfo{author}{\bibfnamefont{V.}~\bibnamefont{Dziom}},
  \bibinfo{author}{\bibfnamefont{A.}~\bibnamefont{Shuvaev}},
  \bibinfo{author}{\bibfnamefont{A.}~\bibnamefont{Pimenov}},
  \bibinfo{author}{\bibfnamefont{M.}~\bibnamefont{Schiebl}},
  \bibinfo{author}{\bibfnamefont{A.~A.} \bibnamefont{Mukhin}},
  \bibinfo{author}{\bibfnamefont{V.~Y.} \bibnamefont{Ivanov}},
  \bibinfo{author}{\bibfnamefont{I.~A.} \bibnamefont{Gudim}},
  \bibinfo{author}{\bibfnamefont{L.~N.} \bibnamefont{Bezmaternykh}},
  \bibnamefont{and} \bibinfo{author}{\bibfnamefont{A.}~\bibnamefont{Pimenov}},
  \bibinfo{journal}{Phys. Rev. B} \textbf{\bibinfo{volume}{92}},
  \bibinfo{pages}{184409} (\bibinfo{year}{2015}),
  \urlprefix\url{https://link.aps.org/doi/10.1103/PhysRevB.92.184409}.

\bibitem[{\citenamefont{Takahashi et~al.}(2016)\citenamefont{Takahashi,
  Kibayashi, Kaneko, and Tokura}}]{takahashi:180404}
\bibinfo{author}{\bibfnamefont{Y.}~\bibnamefont{Takahashi}},
  \bibinfo{author}{\bibfnamefont{S.}~\bibnamefont{Kibayashi}},
  \bibinfo{author}{\bibfnamefont{Y.}~\bibnamefont{Kaneko}}, \bibnamefont{and}
  \bibinfo{author}{\bibfnamefont{Y.}~\bibnamefont{Tokura}},
  \bibinfo{journal}{Phys. Rev. B} \textbf{\bibinfo{volume}{93}},
  \bibinfo{pages}{180404} (\bibinfo{year}{2016}),
  \urlprefix\url{https://link.aps.org/doi/10.1103/PhysRevB.93.180404}.

\bibitem[{\citenamefont{Narita et~al.}(2016)\citenamefont{Narita, Tokunaga,
  Kikkawa, Taguchi, Tokura, and Takahashi}}]{narita:094433}
\bibinfo{author}{\bibfnamefont{H.}~\bibnamefont{Narita}},
  \bibinfo{author}{\bibfnamefont{Y.}~\bibnamefont{Tokunaga}},
  \bibinfo{author}{\bibfnamefont{A.}~\bibnamefont{Kikkawa}},
  \bibinfo{author}{\bibfnamefont{Y.}~\bibnamefont{Taguchi}},
  \bibinfo{author}{\bibfnamefont{Y.}~\bibnamefont{Tokura}}, \bibnamefont{and}
  \bibinfo{author}{\bibfnamefont{Y.}~\bibnamefont{Takahashi}},
  \bibinfo{journal}{Phys. Rev. B} \textbf{\bibinfo{volume}{94}},
  \bibinfo{pages}{094433} (\bibinfo{year}{2016}),
  \urlprefix\url{https://link.aps.org/doi/10.1103/PhysRevB.94.094433}.

\bibitem[{\citenamefont{Masuda et~al.}(2017)\citenamefont{Masuda, Kaneko,
  Yamasaki, Tokura, and Takahashi}}]{masuda:041117}
\bibinfo{author}{\bibfnamefont{R.}~\bibnamefont{Masuda}},
  \bibinfo{author}{\bibfnamefont{Y.}~\bibnamefont{Kaneko}},
  \bibinfo{author}{\bibfnamefont{Y.}~\bibnamefont{Yamasaki}},
  \bibinfo{author}{\bibfnamefont{Y.}~\bibnamefont{Tokura}}, \bibnamefont{and}
  \bibinfo{author}{\bibfnamefont{Y.}~\bibnamefont{Takahashi}},
  \bibinfo{journal}{Phys. Rev. B} \textbf{\bibinfo{volume}{96}},
  \bibinfo{pages}{041117} (\bibinfo{year}{2017}),
  \urlprefix\url{https://link.aps.org/doi/10.1103/PhysRevB.96.041117}.

\bibitem[{\citenamefont{K\'ezsm\'arki et~al.}(2015)\citenamefont{K\'ezsm\'arki,
  Nagel, Bord\'acs, Fishman, Lee, Yi, Cheong, and R\~o\ om}}]{kezsmarki:127203}
\bibinfo{author}{\bibfnamefont{I.}~\bibnamefont{K\'ezsm\'arki}},
  \bibinfo{author}{\bibfnamefont{U.}~\bibnamefont{Nagel}},
  \bibinfo{author}{\bibfnamefont{S.}~\bibnamefont{Bord\'acs}},
  \bibinfo{author}{\bibfnamefont{R.~S.} \bibnamefont{Fishman}},
  \bibinfo{author}{\bibfnamefont{J.~H.} \bibnamefont{Lee}},
  \bibinfo{author}{\bibfnamefont{H.~T.} \bibnamefont{Yi}},
  \bibinfo{author}{\bibfnamefont{S.-W.} \bibnamefont{Cheong}},
  \bibnamefont{and} \bibinfo{author}{\bibfnamefont{T.}~\bibnamefont{R\~o\ om}},
  \bibinfo{journal}{Phys. Rev. Lett.} \textbf{\bibinfo{volume}{115}},
  \bibinfo{pages}{127203} (\bibinfo{year}{2015}),
  \urlprefix\url{https://link.aps.org/doi/10.1103/PhysRevLett.115.127203}.

\bibitem[{\citenamefont{Rikken et~al.}(2002)\citenamefont{Rikken, Strohm, and
  Wyder}}]{rikken:133005}
\bibinfo{author}{\bibfnamefont{G.~L. J.~A.} \bibnamefont{Rikken}},
  \bibinfo{author}{\bibfnamefont{C.}~\bibnamefont{Strohm}}, \bibnamefont{and}
  \bibinfo{author}{\bibfnamefont{P.}~\bibnamefont{Wyder}},
  \bibinfo{journal}{Phys. Rev. Lett.} \textbf{\bibinfo{volume}{89}},
  \bibinfo{pages}{133005} (\bibinfo{year}{2002}),
  \urlprefix\url{https://link.aps.org/doi/10.1103/PhysRevLett.89.133005}.

\bibitem[{\citenamefont{Roth and Rikken}(2000)}]{roth:4478}
\bibinfo{author}{\bibfnamefont{T.}~\bibnamefont{Roth}} \bibnamefont{and}
  \bibinfo{author}{\bibfnamefont{G.~L. J.~A.} \bibnamefont{Rikken}},
  \bibinfo{journal}{Phys. Rev. Lett.} \textbf{\bibinfo{volume}{85}},
  \bibinfo{pages}{4478} (\bibinfo{year}{2000}),
  \urlprefix\url{https://link.aps.org/doi/10.1103/PhysRevLett.85.4478}.

\bibitem[{\citenamefont{Roth and Rikken}(2002)}]{roth:063001}
\bibinfo{author}{\bibfnamefont{T.}~\bibnamefont{Roth}} \bibnamefont{and}
  \bibinfo{author}{\bibfnamefont{G.~L. J.~A.} \bibnamefont{Rikken}},
  \bibinfo{journal}{Phys. Rev. Lett.} \textbf{\bibinfo{volume}{88}},
  \bibinfo{pages}{063001} (\bibinfo{year}{2002}),
  \urlprefix\url{https://link.aps.org/doi/10.1103/PhysRevLett.88.063001}.

\bibitem[{\citenamefont{Wang et~al.}(2015)\citenamefont{Wang, Pascut, Gao,
  Tyson, Haule, Kiryukhin, and Cheong}}]{wang:12268}
\bibinfo{author}{\bibfnamefont{Y.}~\bibnamefont{Wang}},
  \bibinfo{author}{\bibfnamefont{G.~L.} \bibnamefont{Pascut}},
  \bibinfo{author}{\bibfnamefont{B.}~\bibnamefont{Gao}},
  \bibinfo{author}{\bibfnamefont{T.~A.} \bibnamefont{Tyson}},
  \bibinfo{author}{\bibfnamefont{K.}~\bibnamefont{Haule}},
  \bibinfo{author}{\bibfnamefont{V.}~\bibnamefont{Kiryukhin}},
  \bibnamefont{and} \bibinfo{author}{\bibfnamefont{S.-W.}
  \bibnamefont{Cheong}}, \bibinfo{journal}{Sci. Rep.}
  \textbf{\bibinfo{volume}{5}}, \bibinfo{pages}{12268} (\bibinfo{year}{2015}),
  \urlprefix\url{http://dx.doi.org/10.1038/srep12268}.

\bibitem[{\citenamefont{Silwal et~al.}(2013)\citenamefont{Silwal, La-o
  vorakiat, Chia, Kim, and Talbayev}}]{silwal:092116}
\bibinfo{author}{\bibfnamefont{P.}~\bibnamefont{Silwal}},
  \bibinfo{author}{\bibfnamefont{C.}~\bibnamefont{La-o vorakiat}},
  \bibinfo{author}{\bibfnamefont{E.~E.~M.} \bibnamefont{Chia}},
  \bibinfo{author}{\bibfnamefont{D.~H.} \bibnamefont{Kim}}, \bibnamefont{and}
  \bibinfo{author}{\bibfnamefont{D.}~\bibnamefont{Talbayev}},
  \bibinfo{journal}{AIP Advances} \textbf{\bibinfo{volume}{3}},
  \bibinfo{eid}{092116} (\bibinfo{year}{2013}),
  \urlprefix\url{http://scitation.aip.org/content/aip/journal/adva/3/9/10.1063/1.4821548}.

\bibitem[{\citenamefont{McCarroll et~al.}(1957)\citenamefont{McCarroll, Katz,
  and Ward}}]{mccarroll:5410}
\bibinfo{author}{\bibfnamefont{W.~H.} \bibnamefont{McCarroll}},
  \bibinfo{author}{\bibfnamefont{L.}~\bibnamefont{Katz}}, \bibnamefont{and}
  \bibinfo{author}{\bibfnamefont{R.}~\bibnamefont{Ward}},
  \bibinfo{journal}{Journal of the American Chemical Society}
  \textbf{\bibinfo{volume}{79}}, \bibinfo{pages}{5410} (\bibinfo{year}{1957}),
  \eprint{http://dx.doi.org/10.1021/ja01577a021},
  \urlprefix\url{http://dx.doi.org/10.1021/ja01577a021}.

\bibitem[{\citenamefont{Kurumaji et~al.}(2015)\citenamefont{Kurumaji, Ishiwata,
  and Tokura}}]{kurumaji:031034}
\bibinfo{author}{\bibfnamefont{T.}~\bibnamefont{Kurumaji}},
  \bibinfo{author}{\bibfnamefont{S.}~\bibnamefont{Ishiwata}}, \bibnamefont{and}
  \bibinfo{author}{\bibfnamefont{Y.}~\bibnamefont{Tokura}},
  \bibinfo{journal}{Phys. Rev. X} \textbf{\bibinfo{volume}{5}},
  \bibinfo{pages}{031034} (\bibinfo{year}{2015}),
  \urlprefix\url{https://link.aps.org/doi/10.1103/PhysRevX.5.031034}.

\bibitem[{\citenamefont{Kurumaji
  et~al.}(2017{\natexlab{a}})\citenamefont{Kurumaji, Takahashi, Fujioka,
  Masuda, Shishikura, Ishiwata, and Tokura}}]{kurumaji:020405}
\bibinfo{author}{\bibfnamefont{T.}~\bibnamefont{Kurumaji}},
  \bibinfo{author}{\bibfnamefont{Y.}~\bibnamefont{Takahashi}},
  \bibinfo{author}{\bibfnamefont{J.}~\bibnamefont{Fujioka}},
  \bibinfo{author}{\bibfnamefont{R.}~\bibnamefont{Masuda}},
  \bibinfo{author}{\bibfnamefont{H.}~\bibnamefont{Shishikura}},
  \bibinfo{author}{\bibfnamefont{S.}~\bibnamefont{Ishiwata}}, \bibnamefont{and}
  \bibinfo{author}{\bibfnamefont{Y.}~\bibnamefont{Tokura}},
  \bibinfo{journal}{Phys. Rev. B} \textbf{\bibinfo{volume}{95}},
  \bibinfo{pages}{020405} (\bibinfo{year}{2017}{\natexlab{a}}),
  \urlprefix\url{https://link.aps.org/doi/10.1103/PhysRevB.95.020405}.

\bibitem[{\citenamefont{Bertrand and Kerner-Czeskleba}(1975)}]{bertrand:379}
\bibinfo{author}{\bibfnamefont{D.}~\bibnamefont{Bertrand}} \bibnamefont{and}
  \bibinfo{author}{\bibfnamefont{H.}~\bibnamefont{Kerner-Czeskleba}},
  \bibinfo{journal}{J. de Physique} \textbf{\bibinfo{volume}{36}},
  \bibinfo{pages}{379} (\bibinfo{year}{1975}).

\bibitem[{sm:()}]{sm:odefzmo}
\bibinfo{note}{See Supplemental Material at [URL will be inserted by publisher]
  for magnetic susceptibility and hysteresis measurements.}

\bibitem[{\citenamefont{Kurumaji
  et~al.}(2017{\natexlab{b}})\citenamefont{Kurumaji, Takahashi, Fujioka,
  Masuda, Shishikura, Ishiwata, and Tokura}}]{kurumaji:077206}
\bibinfo{author}{\bibfnamefont{T.}~\bibnamefont{Kurumaji}},
  \bibinfo{author}{\bibfnamefont{Y.}~\bibnamefont{Takahashi}},
  \bibinfo{author}{\bibfnamefont{J.}~\bibnamefont{Fujioka}},
  \bibinfo{author}{\bibfnamefont{R.}~\bibnamefont{Masuda}},
  \bibinfo{author}{\bibfnamefont{H.}~\bibnamefont{Shishikura}},
  \bibinfo{author}{\bibfnamefont{S.}~\bibnamefont{Ishiwata}}, \bibnamefont{and}
  \bibinfo{author}{\bibfnamefont{Y.}~\bibnamefont{Tokura}},
  \bibinfo{journal}{Phys. Rev. Lett.} \textbf{\bibinfo{volume}{119}},
  \bibinfo{pages}{077206} (\bibinfo{year}{2017}{\natexlab{b}}),
  \urlprefix\url{https://link.aps.org/doi/10.1103/PhysRevLett.119.077206}.

\bibitem[{\citenamefont{Mih\'aly et~al.}(2004)\citenamefont{Mih\'aly, Talbayev,
  Kiss, Zhou, Feh\'er, and J\'anossy}}]{mihaly:024414}
\bibinfo{author}{\bibfnamefont{L.}~\bibnamefont{Mih\'aly}},
  \bibinfo{author}{\bibfnamefont{D.}~\bibnamefont{Talbayev}},
  \bibinfo{author}{\bibfnamefont{L.~F.} \bibnamefont{Kiss}},
  \bibinfo{author}{\bibfnamefont{J.}~\bibnamefont{Zhou}},
  \bibinfo{author}{\bibfnamefont{T.}~\bibnamefont{Feh\'er}}, \bibnamefont{and}
  \bibinfo{author}{\bibfnamefont{A.}~\bibnamefont{J\'anossy}},
  \bibinfo{journal}{Phys. Rev. B} \textbf{\bibinfo{volume}{69}},
  \bibinfo{pages}{024414} (\bibinfo{year}{2004}),
  \urlprefix\url{http://link.aps.org/doi/10.1103/PhysRevB.69.024414}.

\bibitem[{\citenamefont{K\'ezsm\'arki et~al.}(2014)\citenamefont{K\'ezsm\'arki,
  Szaller, Bordács, Kocsis, Tokunaga, Taguchi, Murakawa, Tokura, Engelkamp,
  R\~o\ om et~al.}}]{kezsmarki:3203}
\bibinfo{author}{\bibfnamefont{I.}~\bibnamefont{K\'ezsm\'arki}},
  \bibinfo{author}{\bibfnamefont{D.}~\bibnamefont{Szaller}},
  \bibinfo{author}{\bibfnamefont{S.}~\bibnamefont{Bordács}},
  \bibinfo{author}{\bibfnamefont{V.}~\bibnamefont{Kocsis}},
  \bibinfo{author}{\bibfnamefont{Y.}~\bibnamefont{Tokunaga}},
  \bibinfo{author}{\bibfnamefont{Y.}~\bibnamefont{Taguchi}},
  \bibinfo{author}{\bibfnamefont{H.}~\bibnamefont{Murakawa}},
  \bibinfo{author}{\bibfnamefont{Y.}~\bibnamefont{Tokura}},
  \bibinfo{author}{\bibfnamefont{H.}~\bibnamefont{Engelkamp}},
  \bibinfo{author}{\bibfnamefont{T.}~\bibnamefont{R\~o\ om}},
  \bibnamefont{et~al.}, \bibinfo{journal}{Nat. Comm.}
  \textbf{\bibinfo{volume}{5}}, \bibinfo{pages}{3203} (\bibinfo{year}{2014}),
  \urlprefix\url{http://dx.doi.org/10.1038/ncomms4203}.

\bibitem[{\citenamefont{Miyahara and Furukawa}(2011)}]{miyahara:073708}
\bibinfo{author}{\bibfnamefont{S.}~\bibnamefont{Miyahara}} \bibnamefont{and}
  \bibinfo{author}{\bibfnamefont{N.}~\bibnamefont{Furukawa}},
  \bibinfo{journal}{Journal of the Physical Society of Japan}
  \textbf{\bibinfo{volume}{80}}, \bibinfo{pages}{073708}
  (\bibinfo{year}{2011}), \eprint{http://dx.doi.org/10.1143/JPSJ.80.073708},
  \urlprefix\url{http://dx.doi.org/10.1143/JPSJ.80.073708}.

\bibitem[{\citenamefont{Miyahara and Furukawa}(2014)}]{miyahara:195145}
\bibinfo{author}{\bibfnamefont{S.}~\bibnamefont{Miyahara}} \bibnamefont{and}
  \bibinfo{author}{\bibfnamefont{N.}~\bibnamefont{Furukawa}},
  \bibinfo{journal}{Phys. Rev. B} \textbf{\bibinfo{volume}{89}},
  \bibinfo{pages}{195145} (\bibinfo{year}{2014}),
  \urlprefix\url{https://link.aps.org/doi/10.1103/PhysRevB.89.195145}.

\bibitem[{\citenamefont{Jia et~al.}(2006)\citenamefont{Jia, Onoda, Nagaosa, and
  Han}}]{jia:224444}
\bibinfo{author}{\bibfnamefont{C.}~\bibnamefont{Jia}},
  \bibinfo{author}{\bibfnamefont{S.}~\bibnamefont{Onoda}},
  \bibinfo{author}{\bibfnamefont{N.}~\bibnamefont{Nagaosa}}, \bibnamefont{and}
  \bibinfo{author}{\bibfnamefont{J.~H.} \bibnamefont{Han}},
  \bibinfo{journal}{Phys. Rev. B} \textbf{\bibinfo{volume}{74}},
  \bibinfo{pages}{224444} (\bibinfo{year}{2006}),
  \urlprefix\url{https://link.aps.org/doi/10.1103/PhysRevB.74.224444}.

\bibitem[{\citenamefont{Arima}(2007)}]{arima:073702}
\bibinfo{author}{\bibfnamefont{T.}~\bibnamefont{Arima}},
  \bibinfo{journal}{Journal of the Physical Society of Japan}
  \textbf{\bibinfo{volume}{76}}, \bibinfo{pages}{073702}
  (\bibinfo{year}{2007}), \eprint{http://dx.doi.org/10.1143/JPSJ.76.073702},
  \urlprefix\url{http://dx.doi.org/10.1143/JPSJ.76.073702}.

\end{thebibliography}

\end{document}